\documentclass[reqno]{article}
\usepackage{alltt}
\usepackage{graphicx}
\title{The non-Abelian field current of the self-interacting quantum electron}
\author{Peter Leifer}
\date{Or-Aqiva, Israel}
\begin{document}
\maketitle
\begin{abstract}
Internal degrees of freedoms of the quantum electron (spin and charge) introduced by Dirac lead to the non-Abelian field configuration of the electron in the complex projective Hilbert space $CP(3)$ of the unlocated quantum states (UQS). Such fields represented by the coefficient functions of the local dynamical variables (LDV's) corresponding $SU(4)$ generators of the Poincar\'e group. These generators describe the deformation of the UQS by the dynamical shifts, boosts and rotations. Interaction of this non-Abelian field with the electrodynamics-like gauge field (internal+external) will suppress the divergency of the Jacobi vector field in the vicinity of the ```north pole" in $CP(3)$. Thereby, the stable ``bundle" of the nearby geodesics comprises the lump-like quantum self-interacting electron.
\end{abstract}
\vskip 0.1cm
\vskip 0.1cm

\section{Introduction}
The structure of the self-interacting quantum electron is in the focus of present work.
In the framework of the Quantum Relativity \cite{Le18/1,Le18/2,Le18/3} I will discuss a new kind of the gauge theory of the extended quantum electron.
I try find some deviations from the Maxwell equations for the single extended quantum electron. De Broglie internal periodic motions if the electron may be related to the spin/charge motions along closed geodesic of the $CP(3)$.

The mass-shell restriction dictates the Clifford algebra for the matrices of Dirac belonging to the $AlgSU(4)$ and the plane wave solution of the Dirac equation for free electron leads to the well known eigenvalue problem \cite{Bethe}. In such a formulation the ``on-shell" condition takes the form of a solvability of the homogeneous linear system $D=(E^2 -m^2c^4-c^2|\vec{p}|^2)^2 = 0$.
These four solutions
\begin{eqnarray}
u_1=1, u_2=0, u_3=\frac{c p_z}{mc^2+\sqrt{m^2c^4+c^2|\vec{p}|^2}}, u_4=\frac{c(p_x+ip_y)}{mc^2+\sqrt{m^2c^4+c^2|\vec{p}|^2}}, \cr
u_1=0, u_2=1, u_3=\frac{c(p_x-ip_y)}{mc^2+\sqrt{m^2c^4+c^2|\vec{p}|^2}}, u_4=\frac{-c p_z}{mc^2+\sqrt{m^2c^4+c^2|\vec{p}|^2}},\cr
u_1=\frac{-c p_z}{mc^2+\sqrt{m^2c^4+c^2|\vec{p}|^2}}, u_2=\frac{-c(p_x+ip_y)}{mc^2+\sqrt{m^2c^4+c^2|\vec{p}|^2}}, u_3=1, u_4=0, \cr
u_1=\frac{-c(p_x-ip_y)}{mc^2+\sqrt{m^2c^4+c^2|\vec{p}|^2}}, u_2=\frac{c p_z}{mc^2+\sqrt{m^2c^4+c^2|\vec{p}|^2}}, u_3=0, u_4=1.
\end{eqnarray}
may be rewritten in inhomogeneous coordinates $\pi^i_{(j)}=\frac{u_i}{u_j}$ in the four maps
$U_1:\{u_1 \neq 0\},U_2:\{u_2 \neq 0\},U_3:\{u_3 \neq 0\},U_4:\{u_4 \neq 0\} $.
Say, solution in the map $U_4:\{u_4 \neq 0\}$ will be as follows
\begin{eqnarray}
\pi^1=\frac{-c(p_x-ip_y)}{mc^2+\sqrt{m^2c^4+c^2|\vec{p}|^2}}; \cr
\pi^2=\frac{c p_z}{mc^2+\sqrt{m^2c^4+c^2|\vec{p}|^2}};\cr
\pi^3=0.
\end{eqnarray}
If one decides to use the local inhomogeneous coordinates $\pi^i$ of the state vector initially, then the single-value solutions of the tree linear inhomogeneous equation may be obtained in Cramer's rule under the condition $D=(E^2 -m^2c^4-c^2|\vec{p}|^2 )^2\neq 0$ \cite{Le07,Le11}. Thereby, the ``off-shell" zone will be accessible for the internal field dynamics of the electron.

One may proof that the the rays of the eigenstates of the Dirac equation for the free electron lie on the geodesic in $CP(3)$  and furthermore simple quadratic functions of these local coordinates lie on the local ``light cone".
Let rewrite (2) in the identical form
\begin{eqnarray}
\pi^1=\frac{-c(p_x-ip_y)}{mc^2+\sqrt{m^2c^4+c^2|\vec{p}|^2}}\frac{c|\vec{p}|}{c|\vec{p}|}; \cr
\pi^2=\frac{c p_z}{mc^2+\sqrt{m^2c^4+c^2|\vec{p}|^2}}\frac{c|\vec{p}|}{c|\vec{p}|};\cr
\pi^3=0.
\end{eqnarray}
and then introducing $\tan(\theta)= \frac{c|\vec{p}|}{mc^2+\sqrt{m^2c^4+c^2|\vec{p}|^2}}$
and $f^1=-(p_x-ip_y), f^2= p_z, f^3=0$ with
$g^2=|f^1|^2+|f^2|^2+|f^3|^2=|\vec{p}|^2=p_x^2+p_y^2+p_z^2$,
one may rewrite the local coordinates of the UQS of the geodesic in $CP(3)$ as follows
\begin{eqnarray}
\pi^1(\theta)=\frac{f^1}{g}\tan(\theta); \cr
\pi^2(\theta)=\frac{f^2}{g}\tan(\theta);\cr
\pi^3=0.
\end{eqnarray}
One may note that the mass $m$ as a parameter may be deleted from the $\tan(\theta)$.
Taking into account that (4) limits $CP(3)$ to $CP(2)$ one may introduce the four variables \begin{eqnarray}
X_0= |\pi^1|^2 + |\pi^2|^2;\cr
X_1= \pi^{1*} \pi^2 + \pi^{2*} \pi^1;\cr
X_2= \frac{1}{i}(\pi^{1*} \pi^2 - \pi^{2*} \pi^1);\cr
X_3=|\pi^1|^2 - |\pi^2|^2;\cr
\end{eqnarray}
belonging to the ``light cone"
$X_0^2-X_1^2-X_2^2-X_3^2=0$. Such ``light cone" is invariant relative finite ``Lorentz transformations". We will use, however, only infinitesimal Poincar\'e generators for the construction of the local DST and non-Abelian field equations of the self-interacting electron.

It is clear that whole geodesic (4) contains the plane waves with the full spectrum of the wave lengths $0\le |\vec{p}| < \infty$. Probably, it is difficult to show that all geodesics rotated by the $H=U(1) \times U(3)$ contains all directions of the $\vec{p}$ but the ``bundle" of nearby geodesics in the vicinity of the basic geodesic (4) will contains the plane waves with small deviations around $\vec{p}$.

Therefore, the main idea is to replace the wave packet of the plane waves that is unstable by the stable ``bundle" of the close geodesics in $CP(3)$. This ``bundle" is shaped by the fields of the geodesic variations of the two types: transversal part generated by the isotropy group $H=U(1)\times U(3)$, and the longitudinal part generated by the coset generators $G/H=SU(4)/ S[U(1)\times U(3)]= CP(3)$. The dynamics of the ``bundle" is governing by the ``quantum Newton equation", i.e. Jacobi equation. Dynamical transition from one UQS on the geodesic to another will be generated by the affine gauge potential $\Gamma^i_{km}$ and the transversal and longitudinal instabilities will be compensated by the fields coefficients of the LDV's corresponding to the $SU(4)$ generators of the shifts, rotations and boosts under the ``inverse representation" of the Poincar\'e group \cite{Le18/2}. Thereby, the dynamical deformation of the spin/charge UQS opens the real way to the interpretation of the boosts as internal electric field, rotations as the internal magnetic field of the spin, and the shifts as the quantum inertia terms.

\section{Dynamical generators of Poincar\'e}
The \emph{existence} of electron and other quantum particles may be physically provided by the  self-interaction that should lead to stable periodic process a la de Broglie.
Closed geodesics in complex projective Hilbert space $CP(N-1)$ is the simplest and natural possibility to describe such internal gauge invariant motions \cite{Le13,Le15,Le16}.
The coset manifold $G/H_{|\psi>}=SU(N)/S[U(1) \times U(N-1)]=CP(N-1)$ contains locally unitary transformations \emph{deforming} ``initial" quantum state $|\psi>$. This means that $CP(N-1)$ contains physically distinguishable, ``deformed" quantum states. Thereby the unitary transformations from $G=SU(N)$ of the basis in the Hilbert space may be identified with the unitary state-dependent gauge field $U(|\psi>)$ that may be represented by the $N^2-1$ unitary generators as functions of the local projective coordinates $(\pi^1,...,\pi^{N-1})$ \cite{Le13}.

The non-linear representation (realization) of the Poincar\'e group by the generators of $SU(4)$ in $CP(3)$ local projective coordinates makes the difference between ``spacetime coordinates" and ``functions of the spacetime coordinates" simply illusory at least in the attempt to find dynamical structure of the self-interacting quantum electron with its EM-like ``field shell". Namely, the commutators of the dynamical shifts generated by the  LDV's arose from the four matrices of Dirac \cite{Le18/1}
should be compared with the six generators of the boosts and rotations started from the well known definitions in terms of the Dirac matrices \cite{Le18/2}. Let me recall briefly  this construction.

The coefficients of the $SU(4)$ generators will be calculated according to the equation
\begin{equation}
\Phi_{\mu}^i = \lim_{\epsilon \to 0} \epsilon^{-1}
\biggl\{\frac{[\exp(i\epsilon \gamma_{\mu})]_m^i \psi^m}{[\exp(i
\epsilon \gamma_{\mu})]_m^j \psi^m }-\frac{\psi^i}{\psi^j} \biggr\}=
\lim_{\epsilon \to 0} \epsilon^{-1} \{ \pi^i(\epsilon
\gamma_{\mu}) -\pi^i \},
\end{equation}\label{6}
\cite{Le13}
that gives coefficient functions
\begin{eqnarray}
\Phi_{0}^1(\gamma_{t})&=&i(\pi^3-\pi^1 \pi^2), \quad \Phi_{0}^2(\gamma_{t})=i(1-(\pi^2)^2),
\quad \Phi_{0}^3(\gamma_{t})=i(\pi^1-\pi^2 \pi^3); \cr
\Phi_{1}^1(\gamma_{1})&=&-i(\pi^2 -\pi^1 \pi^3),
\quad \Phi_{1}^2(\gamma_{1})=-i(-\pi^1 -\pi^2 \pi^3),
\quad \Phi_{1}^3(\gamma_{1})=-i(-1 -(\pi^3)^2); \cr
\Phi_{2}^1(\gamma_{2})&=&-i(i(\pi^2 +\pi^1 \pi^3)),
\quad \Phi_{2}^2(\gamma_{2})=-i(i(\pi^1 +\pi^2 \pi^3)),
\quad \Phi_{2}^3(\gamma_{2})=-i(i(-1 +(\pi^3)^2)); \cr
\Phi_{3}^1(\gamma_{3})&=&-i(-\pi^3 -\pi^1 \pi^2),
\quad \Phi_{3}^2(\gamma_{3})=-i(-1 -(\pi^2)^2),
\Phi_{3}^3(\gamma_{3})=-i(\pi^1 -\pi^2 \pi^3).
\end{eqnarray}\label{15}
for the local ``spacetime shifts" of the operator
\begin{eqnarray}
\vec{P}_{\mu}=  \Phi^i(P_{\mu}) \frac{\partial }{\partial \pi^i} + c.c.
\end{eqnarray}\label{}
Such choice of the vector fields leads to the ``imaginary" basic in local DST which conserves $4D$ Eucledian geometry along geodesic in $CP(3)$ for real four vectors $(p^0,p^1,p^2,p^3)$ and correspondingly $4D$ pseudo-Eucledian geometry for four vectors $(ip^0,p^1,p^2,p^3)$.

The corresponding coefficient functions of the vector fields of the Lorentz generators is as follows for boosts
\begin{equation}
\Phi^i(B_\alpha) = \lim_{\epsilon \to 0} \epsilon^{-1}
\biggl\{\frac{[\exp(\epsilon B_{\alpha})]_m^i \psi^m}{[\exp(
\epsilon B_{\alpha})]_m^j \psi^m }-\frac{\psi^i}{\psi^j} \biggr\}=
\lim_{\epsilon \to 0} \epsilon^{-1} \{ \pi^i(\epsilon
B_{\alpha}) -\pi^i \},
\end{equation}\label{6}
\begin{eqnarray}
\Phi^1(B_x)=\frac{1}{2}(1-(\pi^1)^2),
\Phi^2(B_x)=\frac{-1}{2}(\pi^3+\pi^1\pi^2),
\Phi^3(B_x)=\frac{-1}{2}(\pi^2+\pi^1\pi^3),\cr
\Phi^1(B_y)=-\frac{i}{2}(1+(\pi^1)^2),
\Phi^2(B_y)=-\frac{i}{2}(\pi^3+\pi^1\pi^2),
\Phi^3(B_y)=\frac{i}{2}(\pi^2-\pi^1\pi^3),\cr
\Phi^1(B_z)=-\pi^1,
\Phi^2(B_z)=-\pi^2,
\Phi^3(B_z)=0,
\end{eqnarray}
and for rotations
\begin{equation}
\Phi^i(R_\alpha) = \lim_{\epsilon \to 0} \epsilon^{-1}
\biggl\{\frac{[\exp(\epsilon R_{\alpha})]_m^i \psi^m}{[\exp(
\epsilon R_{\alpha})]_m^j \psi^m }-\frac{\psi^i}{\psi^j} \biggr\}=
\lim_{\epsilon \to 0} \epsilon^{-1} \{ \pi^i(\epsilon
R_{\alpha}) -\pi^i \},
\end{equation}\label{6}

\begin{eqnarray}
\Phi^1(R_x)=\frac{i}{2}(1-(\pi^1)^2),
\Phi^2(R_x)=\frac{i}{2}(\pi^3-\pi^1\pi^2),
\Phi^3(R_x)=\frac{i}{2}(\pi^2-\pi^1\pi^3),\cr
\Phi^1(R_y)=\frac{1}{2}(1+(\pi^1)^2),
\Phi^2(R_y)=-\frac{1}{2}(\pi^3-\pi^1\pi^2),
\Phi^3(R_y)=\frac{1}{2}(\pi^2+\pi^1\pi^3),\cr
\Phi^1(R_z)=-i\pi^1,
\Phi^2(R_z)=0,
\Phi^3(R_z)=-i\pi^3,
\end{eqnarray}
Then the three generators
\begin{eqnarray}
\vec{B}_{\alpha}=  \Phi^i(B_{\alpha}) \frac{\partial }{\partial \pi^i} + c.c.
\end{eqnarray}\label{}
define the boosts and three generators
\begin{eqnarray}
\vec{R}_{\alpha}=  \Phi^i(R_{\alpha}) \frac{\partial }{\partial \pi^i} + c.c.
\end{eqnarray}\label{}
define the rotations.
The commutators of these vector fields may be found in \cite{Le18/2}.

\section{The non-Abelian field current as the boundary conditions for the ``field shell"} The non-Abelian field equations for the internal current of the electron may be written directly from these commutation relations and may be compared with the Maxwell equations if the dynamical shift will be treated as the differentiation in corresponding direction.
The following equations
\begin{eqnarray}
[P_3[P_2,P_1]]=[P_0[P_2,P_1]]=[P_2[P_3,P_1]]\cr =
=[P_0[P_3,P_1]]=[P_1[P_2,P_3]]=[P_0[P_2,P_3]]=0;
\end{eqnarray}
are more strong than the Yang identity. The analog of the Maxwell equation
\begin{eqnarray}
\frac{\partial E_x}{\partial x}+\frac{\partial E_y}{\partial y}+\frac{\partial E_z}{\partial z} = \rho
\end{eqnarray}
looks like
\begin{eqnarray}
[P_1[P_0,P_1]] + [P_2[[P_0,P_2]]+[P_3[P_0,P_3]]\cr=
(\xi^1 = 12i(\pi^1 \pi^2 - \pi^3) , \xi^2= 12i(-1 + (\pi^2)^2), \xi^3 =12i(-\pi^1 +\pi^2 \pi^3)
\end{eqnarray}
with the vector charge.
The following equations
\begin{eqnarray}
[P_1,R_2]-[P_2,R_1]=(\xi^1=2i(\pi^1 \pi^2 + \pi^3),\xi^2= 2i(1 + (\pi^2)^2),\xi^3 =2i(-\pi^1 +\pi^2 \pi^3)),\cr
[P_3,R_1]-[P_1,R_3]=(\xi^1=2(\pi^2+\pi^1 \pi^3),\xi^2=2(\pi^1+\pi^2 \pi^3),\xi^3 =-2(1- (\pi^3)^2),\cr
[P_2,R_3]-[P_3,R_2]=(\xi^1=2i(\pi^1 \pi^3 - \pi^2),\xi^2= 2i(\pi^1+\pi^2 \pi^3),\xi^3 =2i((1 + (\pi^3)^2))
\end{eqnarray}
are similar to the equation
\begin{eqnarray}
\nabla \times \vec{B}= \vec{J}.
\end{eqnarray}
Thereby, old attempts to identify boosts and rotations with electric and magnetic field
have to be reformulated as the intrinsic non-Abelian field current of the quantum electron. The generators of the internal fields (8),(13),(14) are in involution, hence, according to the theorem of Frobenius, they are quite integrable. These field currents serve as the natural boundary conditions for the ``field shell" equations
\begin{eqnarray}\label{43}
 \frac{\partial P^{\mu}}{\partial x^{\mu}} + P^{\mu} (\frac{\partial \Phi_{\mu}^i}{\partial \pi^i} +
\Gamma^i_{il} \Phi_{\mu}^l) +
\frac{\partial K^{\alpha}}{\partial u^{\alpha}} + K^{\alpha} (\frac{\partial \Phi^i(B_{\alpha})}{\partial \pi^i} +
\Gamma^i_{il} \Phi^l(B_{\alpha})) \cr +
\frac{\partial M^{\alpha}}{\partial {\omega}^{\alpha}} + M^{\alpha} (\frac{\partial \Phi^i(R_{\alpha})}{\partial \pi^i} +
\Gamma^i_{il} \Phi^l(R_{\alpha})) + J^i_{;i} = 0.
\end{eqnarray}
with wide class of the TWS's \cite{Le18/3}. This equation resembles the equation for the so-called quantum potential where the role of such potential plays now the divergency of the Jacobi field $Q = J^i_{;i}$. This topic will be discussed elsewhere.
\section{Dynamical spacetime}
The equation (20) follows from the our requirement of the existence of the stable electron tells that the projection of the trajectory of a single quantum particle onto $CP(3)$ should be a geodesic. Then the speed of the UQS components
\begin{eqnarray}
 T^i= \frac{d \pi^i}{d \tau} =  \frac{c}{\hbar}[P^{\mu} \Phi_{\mu}^i+K^{\alpha}\Phi^i(B_{\alpha})
 +M^{\alpha}\Phi^i(R_{\alpha})+ J^i]
\end{eqnarray}\label{}
should obey the nullification of the covariant derivative in the sense of the Fubini-Study metric
\begin{eqnarray}\label{43}
T^i_{;k}= ( P^{\sigma}\Phi_{\sigma}^i)_{;k} + J^i_{;k}= \frac{\partial P^{\sigma}}{\partial \pi^k}\Phi_{\sigma}^i + P^{\sigma} (\frac{\partial \Phi_{\sigma}^i}{\partial \pi^k}+
\Gamma^i_{kl} \Phi_{\sigma}^l) + J^i_{;k} = 0.
\end{eqnarray}
The Jacobi fields to be taken in the fixed basis \cite{Le18/3}.
Additionally, the equation (21) serves as the characteristics
for the PDE ``Schr\"odinger equation"
\begin{eqnarray}\label{43}
i\hbar \frac{d \Psi(\pi,x_{\mu}, \vec{u},\vec{\omega})}{d\tau}  \cr = [cP^{\mu} \Phi_{\mu}^i+K^{\alpha}\Phi^i(B_{\alpha})
 +M^{\alpha}\Phi^i(R_{\alpha})+ J^i]\frac{\partial \Psi(\pi,x_{\mu}, \vec{u},\vec{\omega})}{\partial \pi^i} + c.c. = 0,
\end{eqnarray}
where the coordinates $(\pi^i, x_{\mu}, \vec{u},\vec{\omega})$ correspond to the shifts, rotations, boosts and gauge parameters of the local DST and  $\tau$ is the \emph{quantum elapsed time counted from the start of the internal motion}. This equation expresses  the conservation of the action for the electron. The calculation of the self-energy of the electron postponed for future work.
It contains the non-Abelian field current interacting with EM-like ``field shell" of the electron contains as some part of the internal energy of electron compensating ``divergency" of the Jacobi field. One may assume that the ``Schr\"odinger equation" with the relativistic Hamiltonian vector field
\begin{eqnarray}
\vec{H}=  c[P^{\mu} \Phi_{\mu}^i +K^{\alpha}\Phi^i(B_{\alpha})
 +M^{\alpha}\Phi^i(R_{\alpha})+ J^i] \frac{\partial }{\partial \pi^i} + c.c.
\end{eqnarray}\label{}
may be used for the eigen-value problem in terms of the PDE for the total wave function
$\Psi(\pi,x_{\mu}, \vec{u},\vec{\omega})$.

The DST arose as an ``inverse representation" of the Poincar\'e group by the motions of the UQS in the $CP(3)$. This means physically that dynamical shifts, rotations and boosts represents the ``field shell" of the electron. Thereby, extended quantum electron represented by the field compensating the unstable Jacobi field due to the ``negative resistance" of the affine gauge potential in the fixed local reference frame. Commonly used the reference frame parallel transported along geodesic eliminates the action of the affine potential. This is an analogous of the local ``freely falling down frame" where gravitation effects does not exists. On the other hand the second derivative of the Jacobi field serves as an non-local (field) analog of the acceleration defined by the curvature of $CP(3)$.

The invariant
\begin{eqnarray}
\Omega^2 = G_{ik*}T^iT^{k*}=G_{ik*} \frac{d \pi^i}{d \tau} \frac{d \pi^{k*}}{d \tau}
\end{eqnarray}
leads to the full local 10D DST which conserves in the subspace $4D$ pseido-Eucledian geometry along geodesic has the state-dependent metric tensor
\begin{eqnarray}\label{43}
g_{\rho \sigma} =G_{ik*} \Phi_{\rho}^i\Phi_{\sigma}^{k*} =
G_{ik*} \Phi_{\mu}^i\Phi_{\nu}^{k*}+ G_{ik*} [\Phi_{\alpha}^i\Phi_{\beta}^{k*}+\Phi_{\gamma}^i\Phi_{\delta}^{k*}]+ ...,
\end{eqnarray}
where $1\leq \rho, \sigma \leq 10$, $0 \leq \mu,\nu \leq 3$, and $1 \leq \alpha, \beta, \gamma,\delta \leq 3$. The first term describe the 4D subspace geometry of the DST and additional terms describe the ``diffusion" of the mass-shell.

\section{Conclusion}
The long living discussion of the foundations of quantum mechanics (QM) smoldering almost 100 years but it continues to flare up from time to time.  The wide spectrum of the points of view in this area has been demonstrated last time: from the alarm of the deep misunderstanding \cite{Oxford_Q,Weinberg_17} to the cradling of the physical community: no problems at all \cite{Baumgarten_18}. I try to find the key at the place where it was lost.

Analysis of the localization problem insists to make attempts of the \emph{intrinsic unification} of quantum principles based on the fundamental concept of quantum amplitudes and the principle of relativity ensures the physical equivalence of any conceivable quantum setup. Realization of such program evokes the necessity of the state-dependent affine gauge field in the state space that acquires reliable physical basis.  Representation of such affine gauge field in dynamical space-time has been applied to the relativistic extended self-interacting Dirac's electron. This approach means that the Yang-Mills arguments about the spacetime coordinate dependence of the gauge unitary rotations should be reversed on the dependence of the dynamical spacetime structure on the gauge transformations of the flexible quantum setup of the internal non-Abelian field  currents.

\end{document}